\documentclass[journal]{IEEEtran}

\usepackage{amsmath}
\usepackage{mathtools}
\usepackage{cite}
\usepackage{mathrsfs}
\usepackage{subfig}
\usepackage{array}
\usepackage{stfloats}
\usepackage{color}
\usepackage{multirow}
\usepackage{hhline}

\usepackage{graphicx}

\graphicspath{{figs/}}

\normalsize

\begin{document}
%
\title{{On Resistive Memories: One Step Row Readout Technique and Sensing Circuitry}}

%
\author{Mohammed~E~Fouda,~Ahmed~M.~Eltawil,~\IEEEmembership{Senior~Member,~IEEE,} and~Fadi~Kurdahi,~\IEEEmembership{Fellow,~IEEE}
\thanks{M. E. Fouda, A. Eltawil and F. Kurdahi are with the Department of Electrical Engineering and Computer Science, University of California--Irvine, Irvine, CA 92697-2625 USA (e-mail:foudam@uci.edu).}
\thanks{This work has been submitted to the IEEE for possible publication.
Copyright may be transferred without notice, after which this version may no longer be accessible}
}

\markboth{}
{Fouda \MakeLowercase{\textit{et al.}}: One Step Row Readout Technique for High-Density Resistive Memories}
%
\maketitle

\begin{abstract} Transistor-based memories are rapidly approaching their maximum density per unit area. Resistive crossbar arrays enable denser memory due to the small size of switching devices. However, due to the resistive nature of these memories, they suffer from current sneak paths complicating the readout procedure. In this paper, we propose a row readout technique with circuitry that can be used to read {selector-less} resistive crossbar based memories. High throughput reading and writing techniques are needed to overcome the memory-wall bottleneck problem and to enable near memory computing paradigm. The proposed technique can read the entire row of dense crossbar arrays in one cycle, unlike previously published techniques. The requirements for the readout circuitry are discussed and satisfied in the proposed circuit. Additionally, an approximated expression for the power consumed while reading the array is derived. A figure of merit is defined and used to compare the proposed approach with existing reading techniques. Finally, a quantitative analysis of the effect of biasing mismatch on the array size is discussed.
\end{abstract}

\begin{IEEEkeywords}
ReRAM, RRAM, Resistive Memories, 3D memories, Sneak Path, Crossbar Arrays
\vspace{-0.1in}
\end{IEEEkeywords}

%
\IEEEpeerreviewmaketitle

\section{Introduction}
\IEEEPARstart{O}{ver} the last decade, emerging nonvolatile memories (NVMs), such as phase change memory (PCRAM), ferroelectric memory (FeRAM), spin transfer torque magnetic memory (STT-MRAM), and resistive memory (RRAM), have shown high potential as alternatives for floating-gate-based nonvolatile memories \cite{daly2017through}. 
RRAMs are considered the best candidate for the next generation nonvolatile memory due to their high reliability, fast access speed, multilevel capabilities and stack-ability creating 3D memory architectures \cite{liu2014130
}.
To achieve higher density memories, switching devices alone are sandwiched between the crossbar metal layers without using access devices such as transistors, diodes and selectors\cite{kawahara20138}. In some cases, the switching devices might have exponential behavior (the selector is embedded inside the switching device) such as FAST selector \cite{jo20143d}. The main drawback of selector-less (gate-less) crossbar-based memories is the sneak path effects which limit the readability of the array. Conventional reading approaches for selector-less crossbars suffer from the sneak path loading which makes reading the data very difficult, and even impossible at times. On the other hand, writing is done through accessing one device at time using one of two bias schemes; either  1/2 bias scheme or 1/3 bias scheme \cite{chen2015analysis}.     

\begin{figure}[!t]
\centering
\vspace{-0.15in}
\subfloat[]{\includegraphics[width=0.48\linewidth]{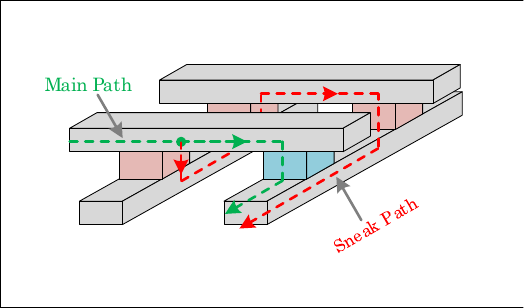}%
\label{Fig2a}}
\hfil
\subfloat[]{\includegraphics[width=0.48\linewidth]{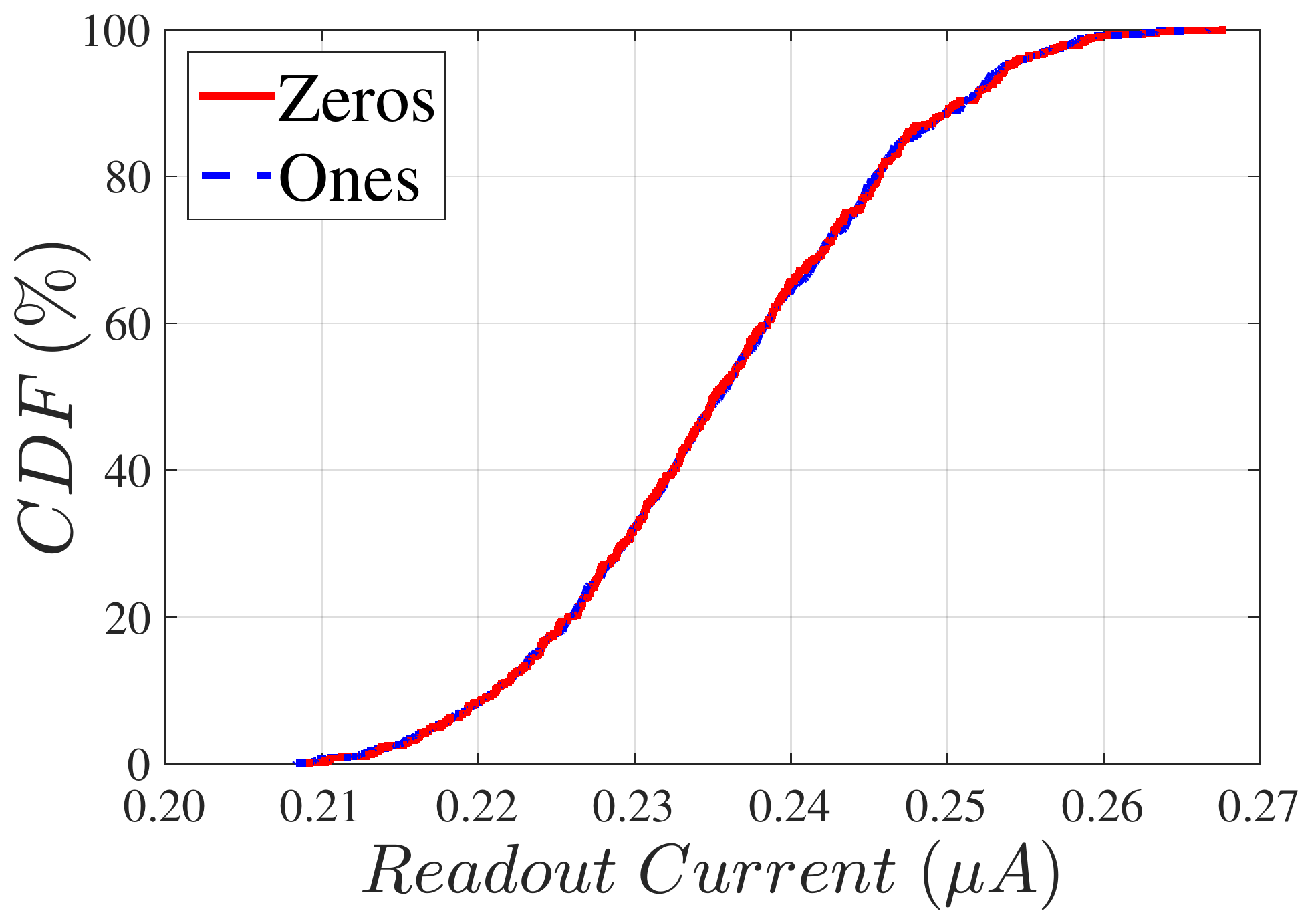}%
\label{Fig2b}}
\vspace{-0.08in}
\caption{(a) Crossbar array with the sneak path problem, and (b) cumulative probability of reading $512\times 512$ array.}
\label{Fig2}
\vspace{-0.22in}
\end{figure}

The sneak paths problem arises because there are many paths from the inputs to the outputs. Figure \ref{Fig2a} shows the sneak path in $2\times2$ crossbar array. The sneak path current is added to the main path current which disturb the cell reading. Figure \ref{Fig2b} shows the cumulative probability of the readout current for $512\times 512$ crossbar array  with $LRS=1M\Omega,\, HRS=1G\Omega$ and $10\Omega$ line resistance. The readout currents corresponding to LRS and HRS are totally overlapped. As a result, it is impossible to find a threshold to differentiate between the two states even with very large switching resistance values.

Recently, different readout techniques were proposed to address this problem in high-density arrays using two main procedures. (a) Reading and writing the stored data multiple times such as \cite{vontobel2009writing,zidan2014memristor} which require an Analog to Digital Converter (ADC), as well as registers and comparators. (b) Dispersing predefined dummy cells in the array to assist in reading the data such as \cite{zidan2016single,manem2010design}. These dummy cells should be initialized first, which requires more than one cycle to read a certain cell and requires locality property, ADC and comparators, all of which limit the applicability of the technique. Other techniques have been proposed to read the data in parallel by adding sensing resistors to the bitlines and sense the voltage across the resistors \cite{liang2010cross} which loads the crossbar and does not mitigate the sneak path effects.{Yet another approach is to keep the bitlines floating and sense the bitlines voltages as discussed in \cite{Fouda2017one}. The floating bitlines schemes can read array sizes up $128\times 128$ without errors. Both resistive load and floating reading schemes are suitable only for small size arrays}. In this brief, a sneak path mitigation readout technique for high density 3D resistive memories is proposed in addition to the required peripheral readout circuitry. 

\vspace{-0.15in}
\section{Proposed Readout Technique}
\begin{figure}[!t]
\vspace{-0.3in}
\centering
\subfloat[]{\includegraphics[width=0.5\linewidth]{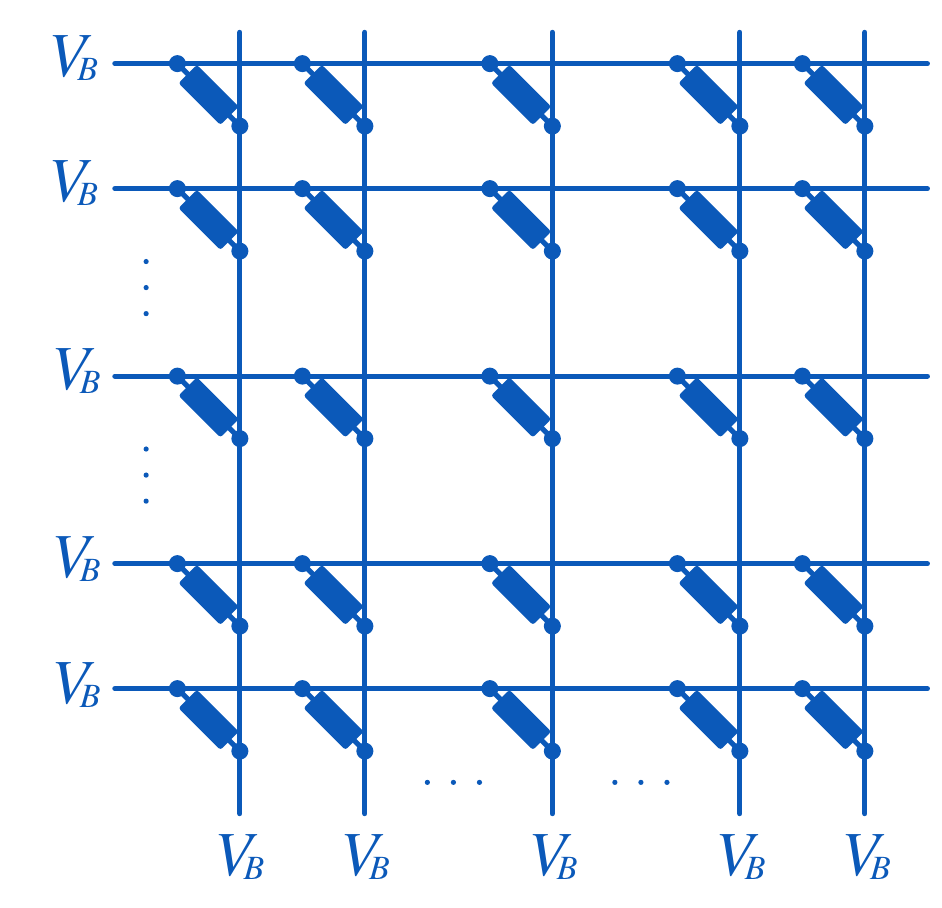}%
\label{Fig3a}}
\hfil
\subfloat[]{\includegraphics[width=0.5\linewidth]{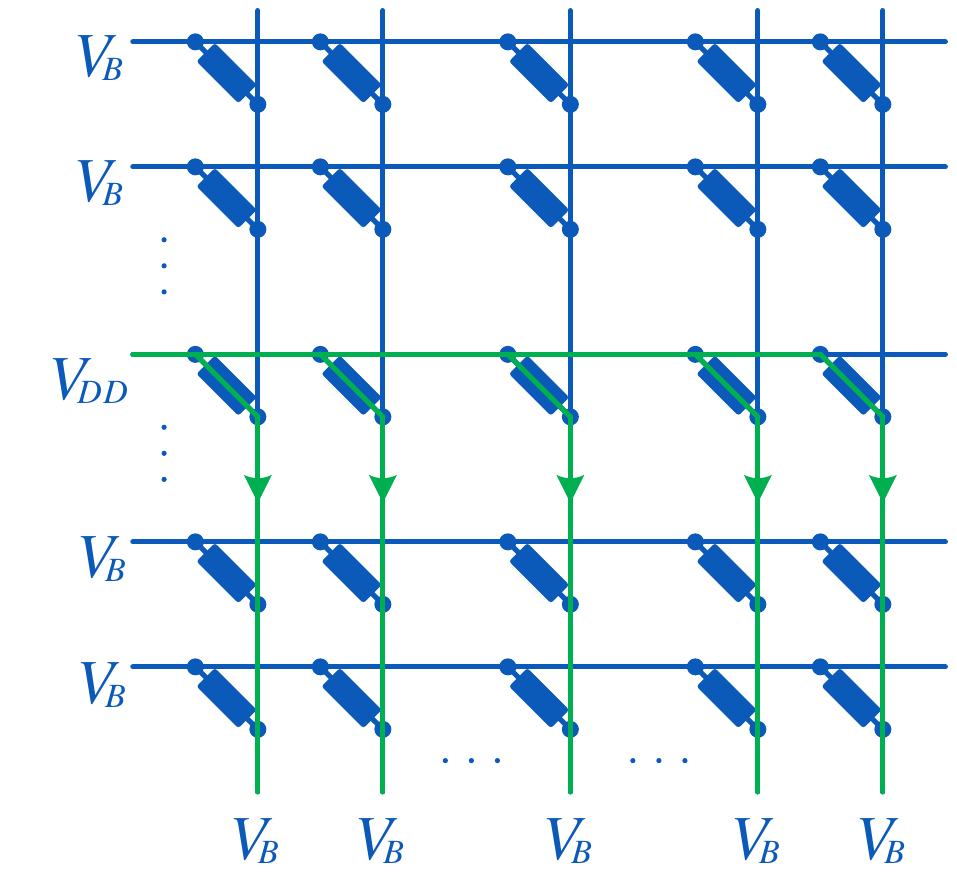}%
\label{Fig3b}}
\vspace{-0.05in}
\caption{(a) Biasing scheme and (b) row reading scheme.}
\label{Fig3}
\vspace{-0.25in}
\end{figure}

Our target is to design circuitry that can be used for reading stacked resistive crossbar arrays. As is clear from previously published techniques, the sneak path problem appears due to the existence of many paths between the input and the outputs. To avoid this problem, we present a simple solution which mitigates the sneak path problem and maximizes throughput. Consider the case, where all the crossbar array input and output ports are biased to a certain bias voltage, $V_B$, as shown in Fig. \ref{Fig3a}. For simplicity, assume that the input and output ports are grounded. Consequently, no current flows across the crossbar array. However, when a $V_{DD}$ signal is applied to one of the input rows, current flows from this input to all the outputs and no current flows across the other rows where the voltage drop across the other rows is zero as shown in Fig. \ref{Fig3b}. The current is absorbed by the sensing circuit. This current is proportional to the resistance of the switching device. In this way, the sneak current paths are eliminated. By reading this current, it is easy to distinguish between the low resistance and high resistance states. The architecture is inherently parallel, where all the row data can be accessed in parallel, enabling high throughput applications.  

\begin{figure}[!t]
\vspace{-0.05in}
\centering
\includegraphics[width=1\linewidth]{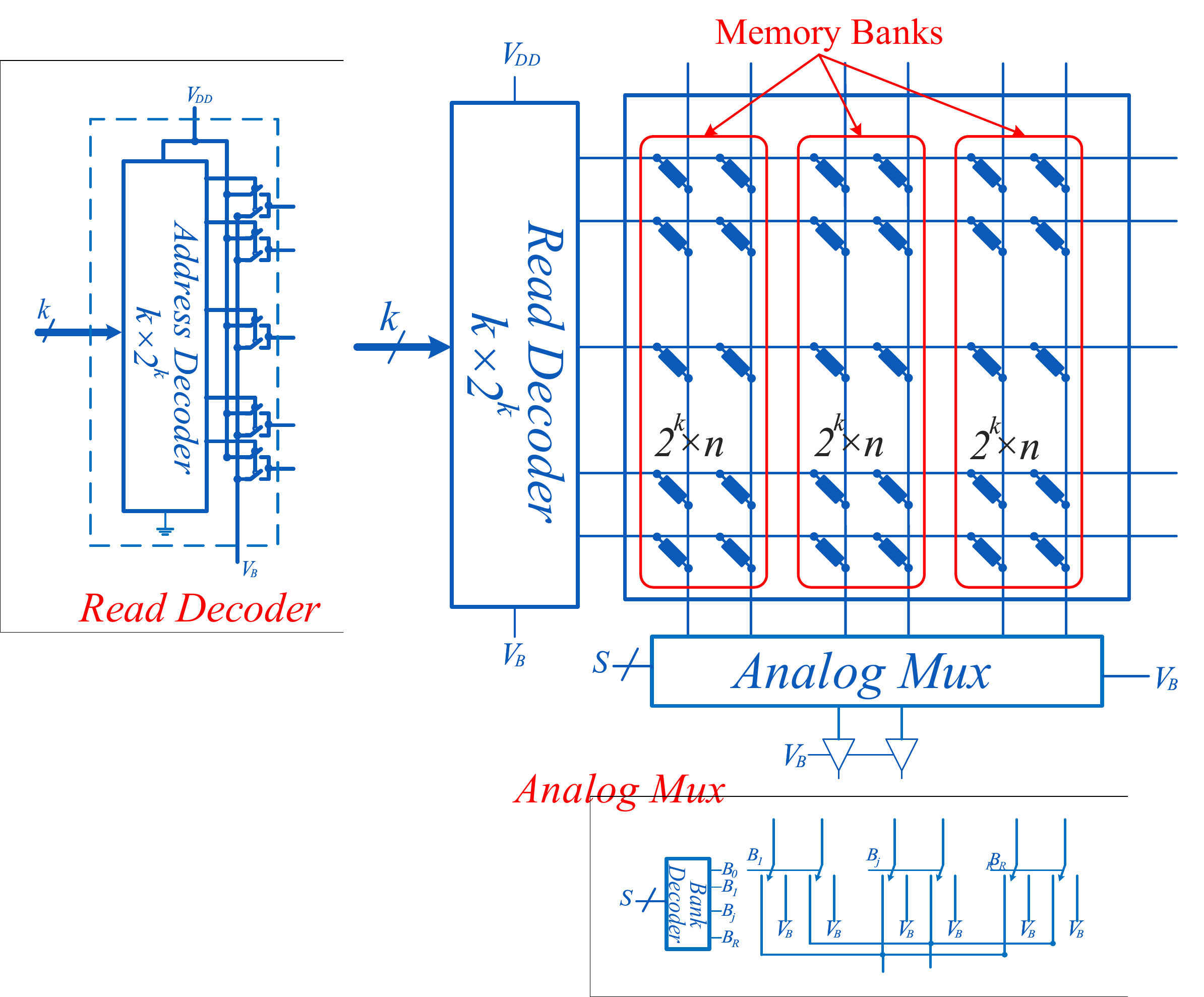}
\caption{Schematic of full single-layered crossbar array.}
\label{Fig4}
\vspace{-0.25in}
\end{figure}

\textit{Memory Architecture:} To read the $(i,j)$ cell, a $V_{DD}$ signal is applied to the $i^{th}$ row and the current of output port number $j$ is sensed {where $1\leq i\leq M$ and $1\leq j\leq N$}. This path can be modeled as resistor, $R_m$, between input and output ports which is either LRS or HRS. The maximum and minimum resistance can be used to determine the sensitivity of sensing circuit as will be discussed later. It is worth noting that this technique can be generalized to any crossbar size since the sneak current that is resultant from multi-paths is mitigated by this reading technique. Figure \ref{Fig4} shows the full schematic of the single layered crossbar array. The crossbar inputs are connected to the read decoder which selects only one input according to the input address. The selected line is biased by $V_{DD}$ and unselected lines are biased to $V_B$. Grounding the lines is a special case of the general bias voltage $V_B$. Furthermore, if smaller word-lengths are needed, banking can be applied as shown in the figure, where each bank has size of $M\times n$ and the total number of banks is $R=N/n$. Consequently, one row per bank is read each clock cycle which means $n$ readout circuits are needed. In order to choose between the banks, an analog mux is necessary which can easily be constructed using switches. Note that the bitlines of all unselected banks should be biased to $V_B$ to guarantee the aforementioned scenario. The outputs of the analog mux are connected directly to the reading circuits which bias the selected bank to $V_B$ and sense the current.

\begin{figure}[!t]
\vspace{-0.05in}
\centering
\subfloat[]{\includegraphics[width=0.50\linewidth,height=0.6\linewidth]{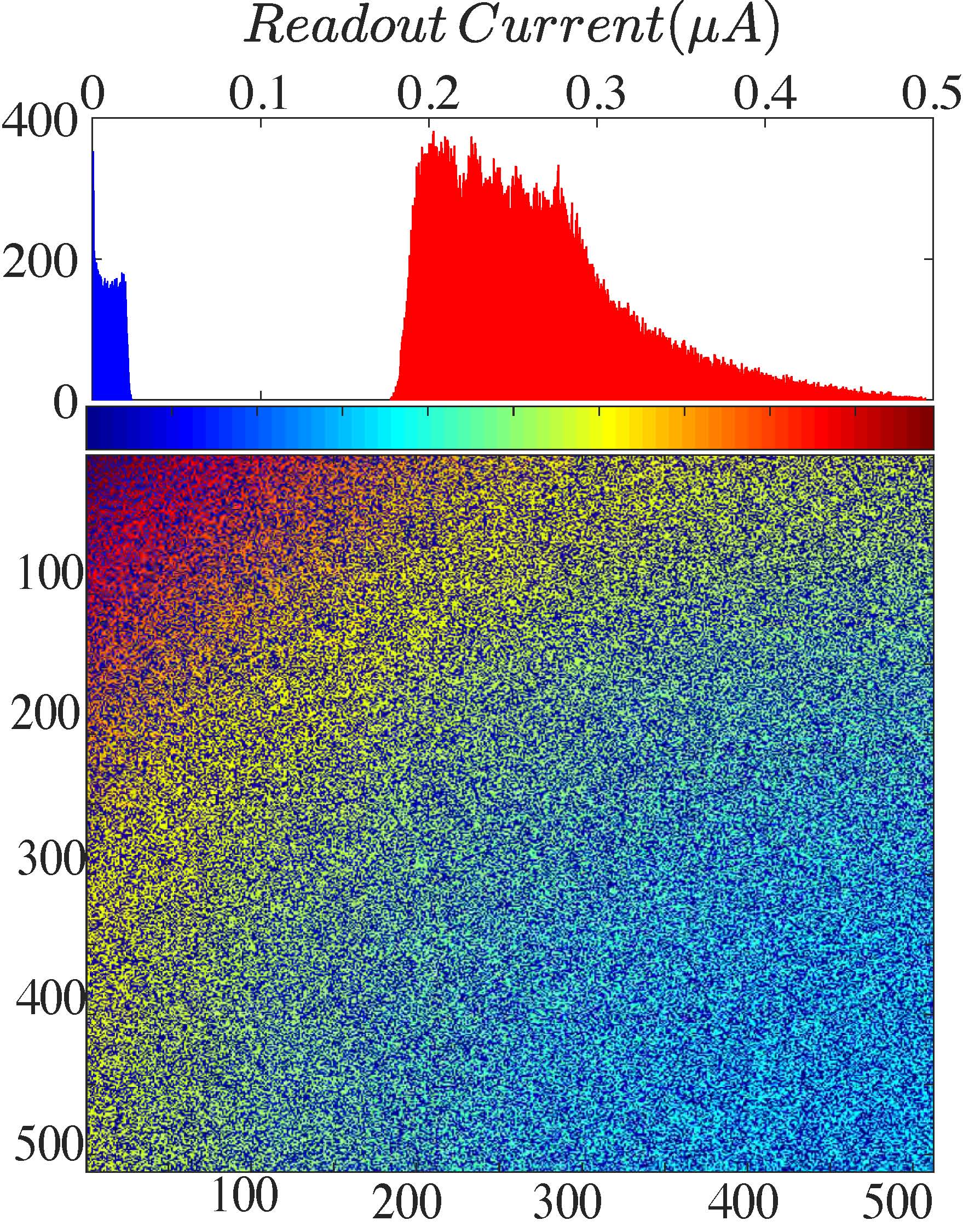}%
\label{Fig5acom}}
\hfil
\subfloat[]{\includegraphics[width=0.50\linewidth,height=0.6\linewidth]{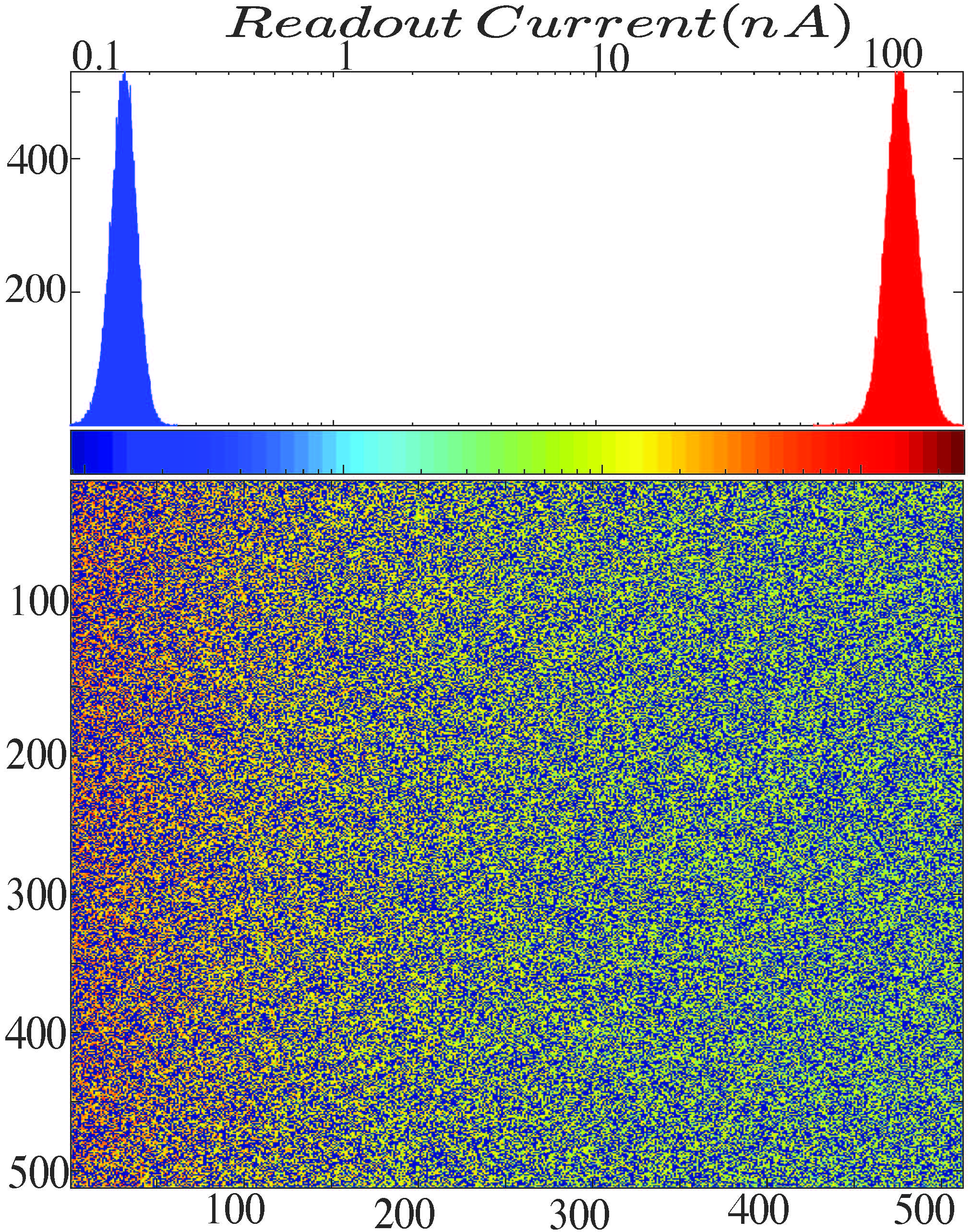}%
\label{Fig5bcom}}
\vspace{-0.1in}
\caption{{ Combined plot for the $512\times 512$ array sensed current density with histogram for (a)linear and (b) nonlinear devices.}}
\label{FigCombined}
\vspace{-0.25in}
\end{figure}

\textit{Effect of Wire Resistance:} Wire resistance is inevitable in such crossbar arrays which effects the reading technique making the voltage across the selected devices less than $V_{DD}-V_B$ by a factor which is a function of the stored data and cell location. Due to the random nature of the data, it is hard to estimate this factor analytically. Figure \ref{FigCombined} shows the sensed current density of each cell in $512\times 512$ array, in addition to the histogram of the sensed current for both linear and nonlinear devices with $10\Omega$ wire resistance { and 10\% resistance variations in both states.} Generally, the wire resistance creates leakage paths from the selected wordline to the unselected wordlines. However, with these input to input leakage paths, the technique is still able to distinguish between LRS and HRS with wide current range for linear devices. On the other hand, the nonlinear devices have exponential voltage-dependency modeled as $I=k\times \sinh(a V)$ where $V$ is the voltage across the resistive device, $k$ and $\alpha$ are the fitting parameters\cite{zidan2016single}. Using such devices improves the sensing current range due to the high resistance facing the leakage currents in the input ports. 

\vspace{-0.15in}
\section{Proposed current sensing circuitry}
The readout circuit to sense the output current of the crossbar should satisfy the following specifications: 

1) the sensing terminal has a fixed bias voltage, $V_B$, and

2) the range of sensed current needs to be identified. 

One way to satisfy these requirements is to use the current conveyor concept where the applied voltage of $port\,1$  is mirrored to $port\,2$, while the input current to $port\,2$  is mirrored to $port\,3$ { as shown in Fig. \ref{Fig5a}}\cite{sedra1990current} {(see supplementary material S1 for more details about current conveyor working principle)}. $Port\,1$ can be biased to $V_B$ which is mirrored to $port\,2$ and the input current to $port\,2$ is around $(V_{DD}-V_{B})/ R_{m,i,j}$ where $R_{m,i,j }$ is the $(i,j)$ cell resistance which is either $LRS$ or $HRS$.  
At this point, it is easy to distinguish between the $LRS$ and $HRS$ by designing suitable readout circuit. 

The continuous behavior of RRAMs enables the ability of storing multi-level data.  The proposed reading technique can be used to read multi-level resistive memories as well since it reads the selected cell resistance only. Thus, a suitable readout circuitry that differentiates between the states is needed. In this work, we focus only on reading binary resistive memories. The proposed circuit is divided into two parts; 1) current sensing circuit which should have the aforementioned specifications { which works as a transimpedance amplifier} and 2) a latched comparator to distinguish between the two states and also to latch the data.
\vspace{-0.15in}
\subsection{The Proposed Current Sensing Circuit}
\begin{figure}[!t]
\vspace{-0.1in}
\centering
\subfloat[]{\includegraphics[width=0.3\linewidth]{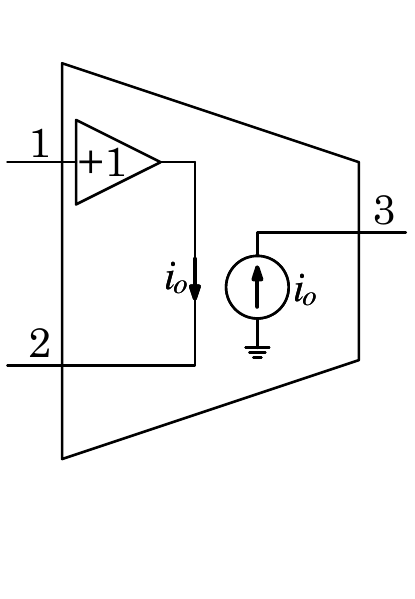}%
\label{Fig5a}}
\hfil
\subfloat[]{\includegraphics[width=0.65\linewidth,height=0.5\linewidth]{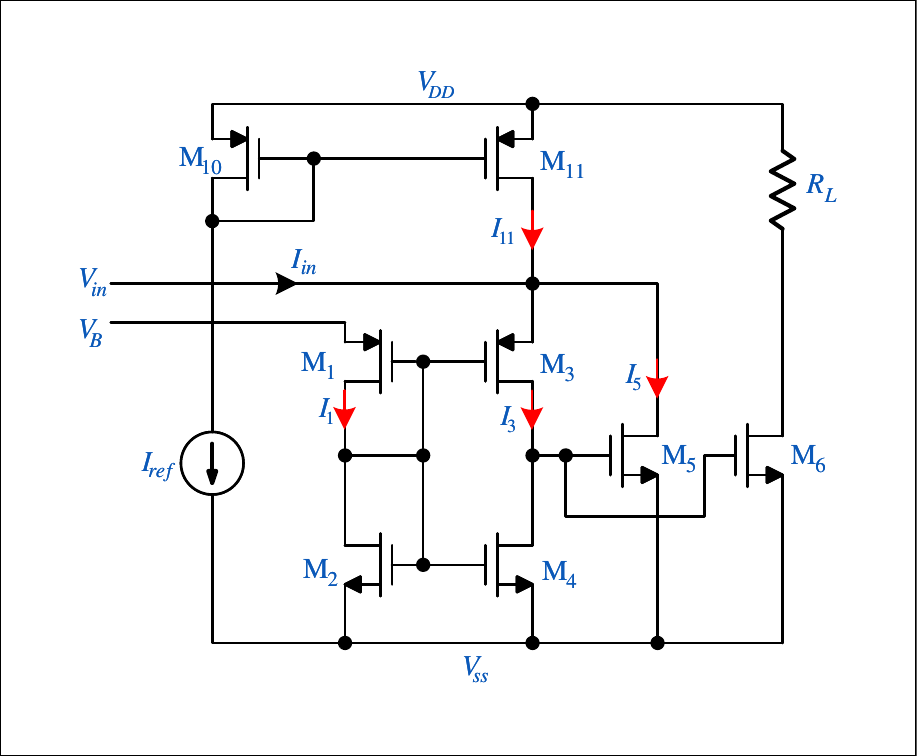}%
\label{Fig5b}}
\vspace{-0.10in}
\caption{(a) Current conveyor principle  and (b) schematic of proposed current sensing circuit.}
\label{Fig5}
\vspace{-0.25in}
\end{figure}

{The proposed circuit is based on the current conveyor concept \cite{sedra1990current,elwan1996cmos} as shown in Fig. \ref{Fig5b}. The voltage $V_B$  is the biasing voltage and can take any value as long as $M_1$ and $M_2$ are kept in the saturation region. $M_2$ and $M_4$ are designed so that the current passing through $M_2$ is mirrored to $M_4$. Consequently, the current passing through $M_1$ is equal to the current passing through $M_3$. Assuming that all transistors are in the saturation region, It can be proven that $V_{in}=V_B$ (supplementary material S3.1). Now, the first condition in the reading circuit is satisfied. 

By applying KCL at the input node, the current passing through $M_5$ is $I_5=I_{11}+I_{in}-I_3$. This current is mirrored through $M_6$ to the output node and imposed into the load resistance creating the output voltage, $ V_o=V_{DD}-(I_{11}+I_{in}-I_3 ) R_L$ where $I_{11}$ is mirrored from $I_{10}$ and its value is $\alpha I_{ref}$ where $\alpha$ is the ratio between the aspect ratios. $I_3$ is a constant current and equals $I_1$ due to the current mirror effect. The analysis for the value of $I_1$ can be found in the supplementary material. Consequently, the output voltage is $V_o=V_{ref}-I_{in} R_L$, where $V_{ref}=V_{DD}-(\alpha I_{ref}-I_1 ) R_L$. The input current $I_{in}$ is either $(V_{DD}-V_B)/LRS$ or $(V_{DD}-V_B)/HRS$. Thus, we have two outputs, $V_{oh}$  and $V_{ol}$ corresponding to HRS and LRS, respectively. It is necessary to widen the difference between $V_{ol}$ and $V_{oh}$ to easily distinguish between the states, which is possible by controlling the values of $R_L$, $\alpha$ and $I_1$. This circuit can be followed by either a buffer or a latch circuit so that the output swings between $V_{DD}$ and $V_{ss}$. To avoid having a large loading resistor occupying a large area, with limited output voltage range, a latched comparator is necessary.}

\vspace{-0.1in}
\subsection{Latched Readout Circuit}
\begin{figure}[!t]
\centering
\includegraphics[width=\linewidth]{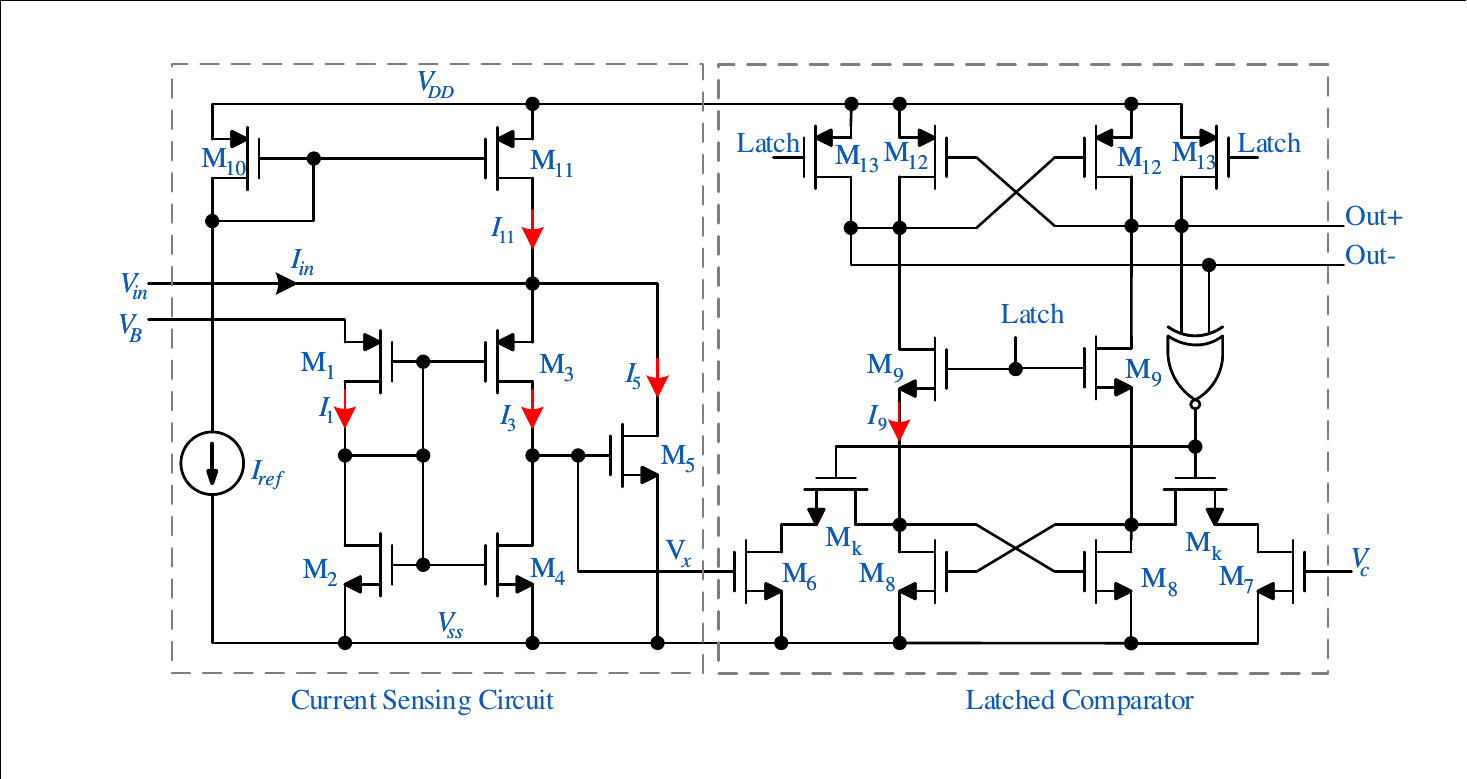}
\vspace{-0.2in}
\caption{Schematic of the full proposed readout circuit.}
\label{ProposedCit}
\vspace{-0.1in}
\end{figure}

\begin{figure}[!t]
\vspace{-0.05in}
\centering
\includegraphics[width=0.75\linewidth,height=0.4\linewidth]{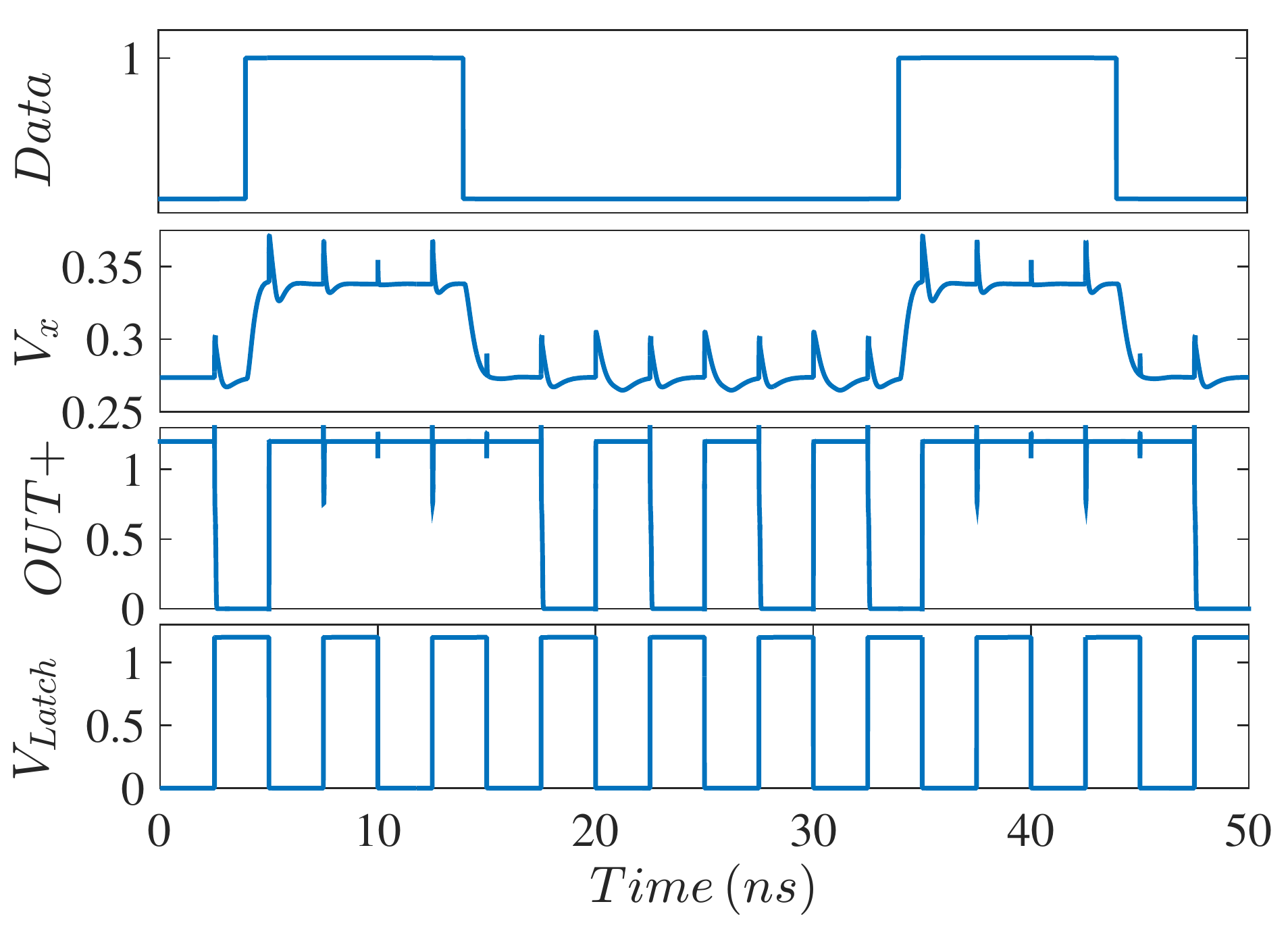}
\vspace{-0.05in}
\caption{Simulation results of the proposed circuit.}
\label{Results}
\vspace{-0.25in}
\end{figure}

Instead of reading the output current from the loading resistor, this current can be connected directly to a latched comparator as shown in Fig. \ref{ProposedCit}. The gate voltage of $M_5$ changes depending on the current passing though it which is mirrored in $M_6$. The mirrored current is compared with the constant current generated by $M_7$ due to constant voltage $V_c$. The $M_k$ transistors are used to reduce kick back noise where the latch signal voltage goes back to the input signal which may alter the data. This latched comparator is introduced and analyzed in \cite{huang2013design}. In the reset case, both outputs are equal to $V_{DD}$ where the output of the XOR gate is zero. On the other hand, in the set case, $V_{Latch}=V_{DD}$. Fig. \ref{Results} illustrates an example of transient simulation of reading random data from a certain column using the proposed circuit. The readout circuit is designed to satisfy the aforementioned conditions (see supplementary material S2.3 for circuit values). The circuit is implemented using TSMC65nm. The area of the entire readout circuitry is about $55 \mu m^2$ with $1.92 \mu W$ total power consumption, at 1.2V supply.
\vspace{-0.1in}
\section{Discussion and Comparison}
\subsection{Power Consumption Estimation}
Power consumption is very critical in resistive memories since they consume a lot of power during the reading and the writing operations due to their inherent resistive nature. Thus, it is essential to estimate the power consumption inside the crossbar array as well as the power density. The device nonlinearity highly affects the power consumption where the higher the nonlinearity, the higher the resistance which implies less power. 

\begin{figure}[!t]
\vspace{-0.1in}
\centering
\subfloat[]{\includegraphics[width=0.50\linewidth,height=0.35\linewidth]{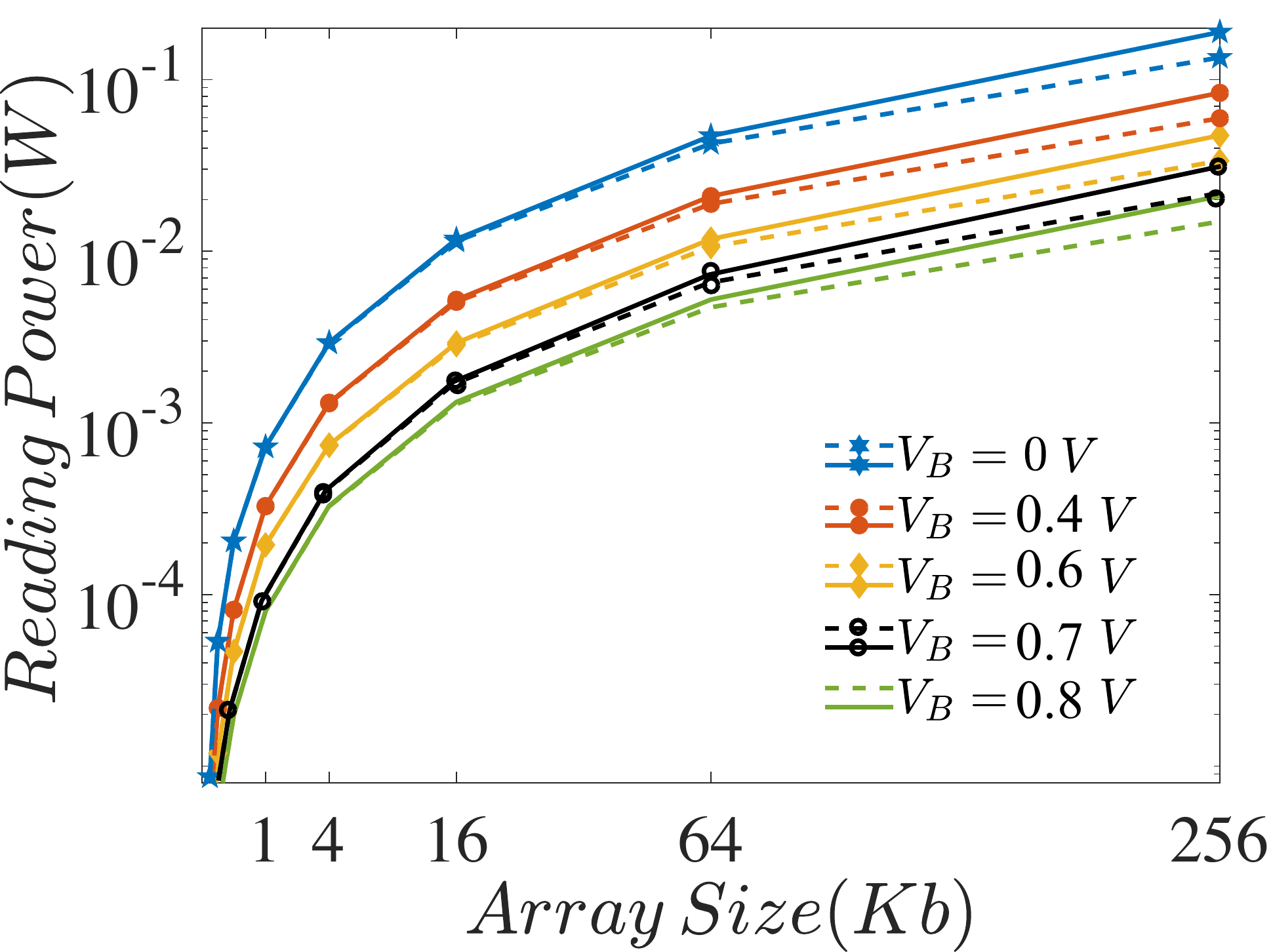}%
\label{PowerCurvea}}
\hfil
\subfloat[]{\includegraphics[width=0.50\linewidth,height=0.35\linewidth]{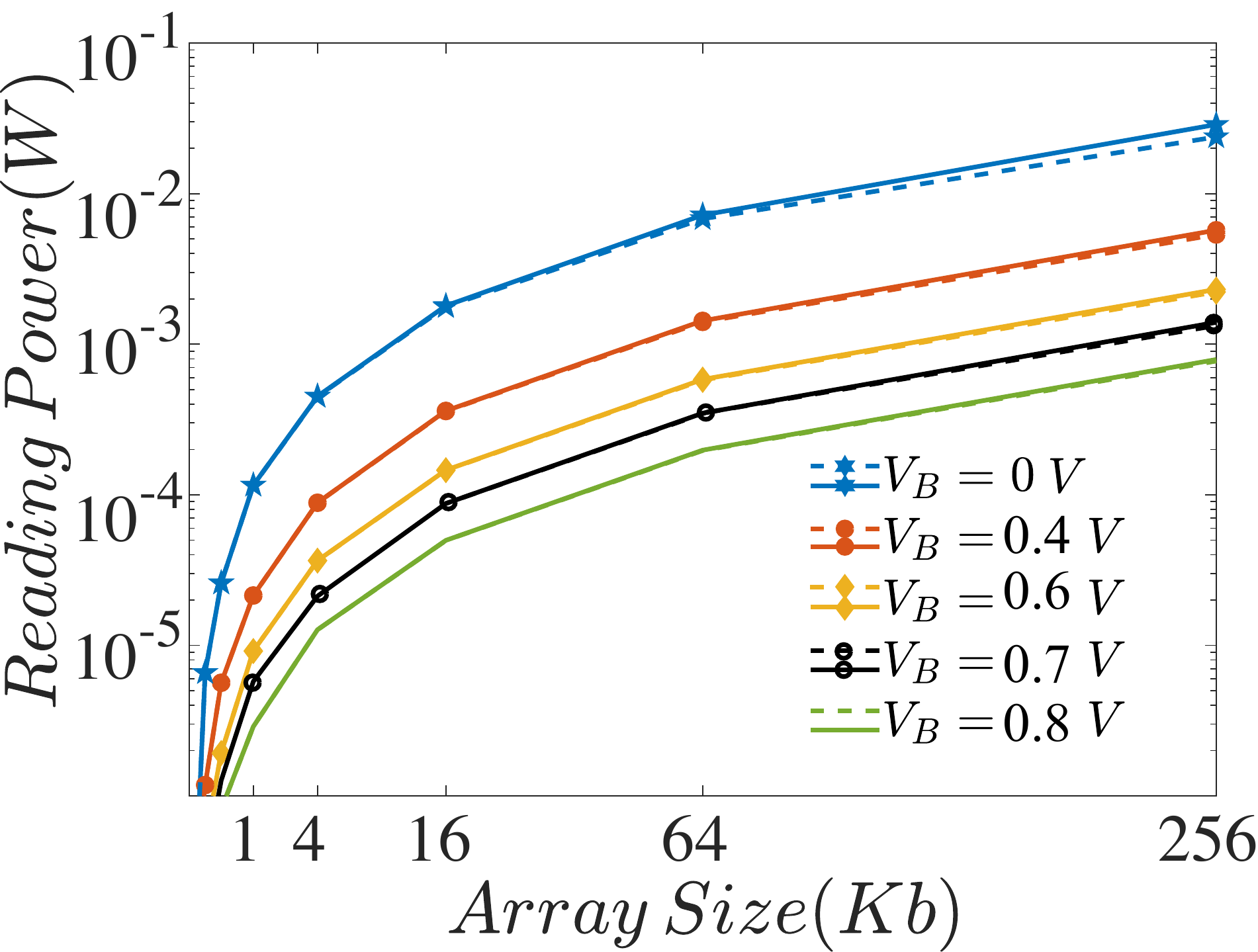}%
\label{PowerCurveb}}
\vspace{-0.1in}
\caption{{Reading power versus the array size at $V_{DD}=1.2V$ with and without wire resistance (dashed lines and solid lines) for (a) linear switching and (b) nonlinear switching devices}}
\label{PowerCurve}
\vspace{-0.2in}
\end{figure}

In case of linear devices, where the resistive device states are constant and not a function of the applied voltage, the power consumption for reading one wordline (for both with and without banking) can be approximated by multiplying the voltage drop times the input current, ignoring wire resistance, $P_{wl_i}=\sum_{j=1}^N \frac{(V_{DD}-V_B)^2}{R_{m,i,j}}$. Thus, the maximum and minimum power consumption are around $M\,N\,R\, (V_{DD}-V_B)^2 /LRS$ and $M\,N\,R\, (V_{DD}-V_B)^2/HRS$, respectively. on the other hand, in case of the nonlinear devices, the voltage across the switching device is still around $V_{DD}-V_B$. Thus, the power consumption per wordline is $P_{wl_i}=\sum_{j=1}^N k_{i,j}(V_{DD}-V_{B}) \sinh(a_{i,j} (V_{DD}-V_B))$ by ignoring wire resistance. Figure \ref{PowerCurve} shows the reading power consumption with and without including the line resistance for different biasing voltages and different array sizes. The figure is plotted for nonlinear switching devices, with $k_{on}, k_{off}$ and $a$ are $1e-8, 1e-11$ and $3$, respectively\cite{zidan2014memristor}. Clearly, by increasing the biasing voltage, $V_B$, a lower power consumption is obtained. However, this reduces the sensing current margin which highly affects the sensing circuit. Therefore, it is important to study the effect of changing $V_B$.

\begin{figure}[!t]
\vspace{-0.15in}
\centering
\subfloat[]{\includegraphics[width=0.5\linewidth]{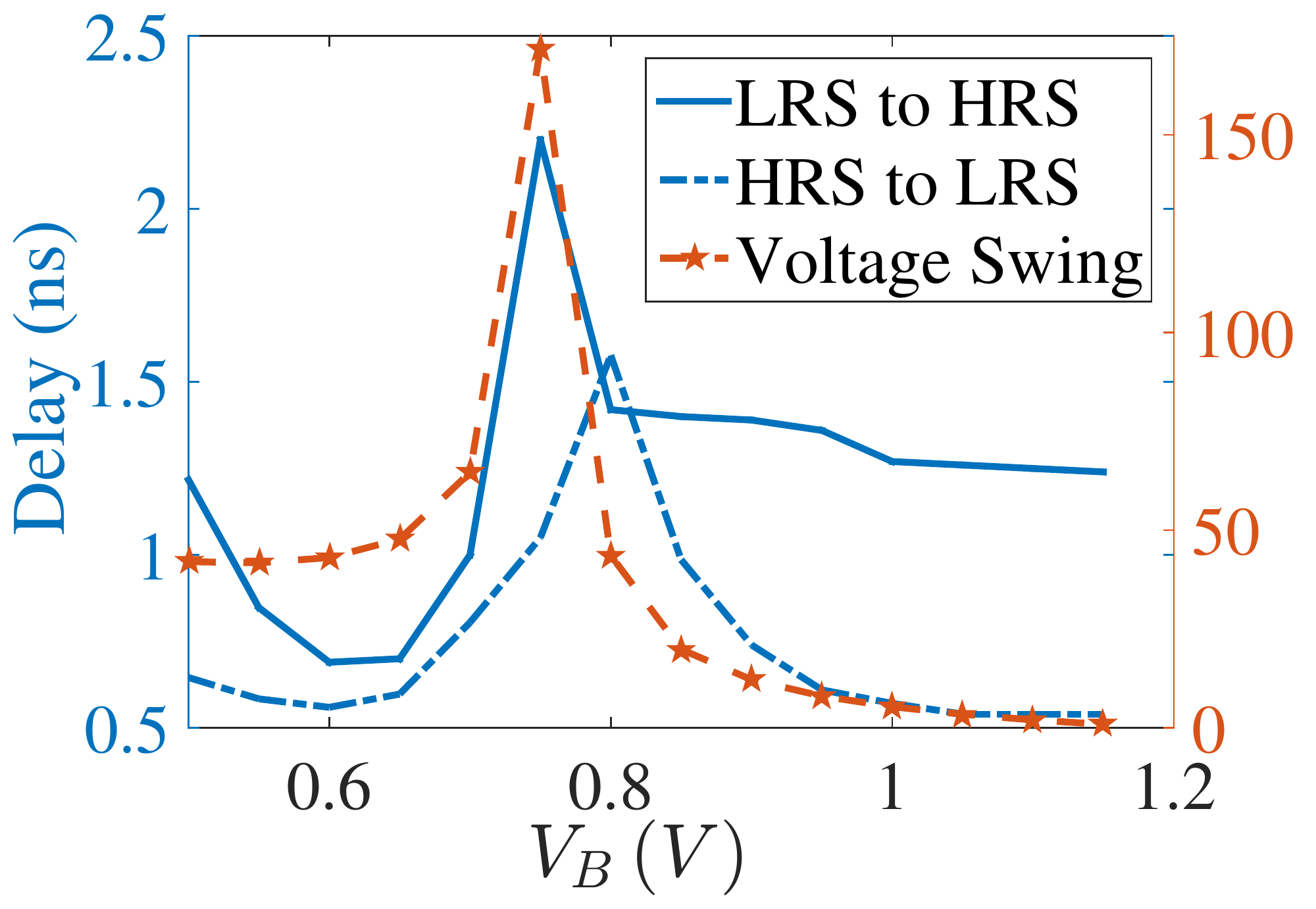}%
\label{Fig10a}}
\hfil
\subfloat[]{\includegraphics[width=0.5\linewidth]{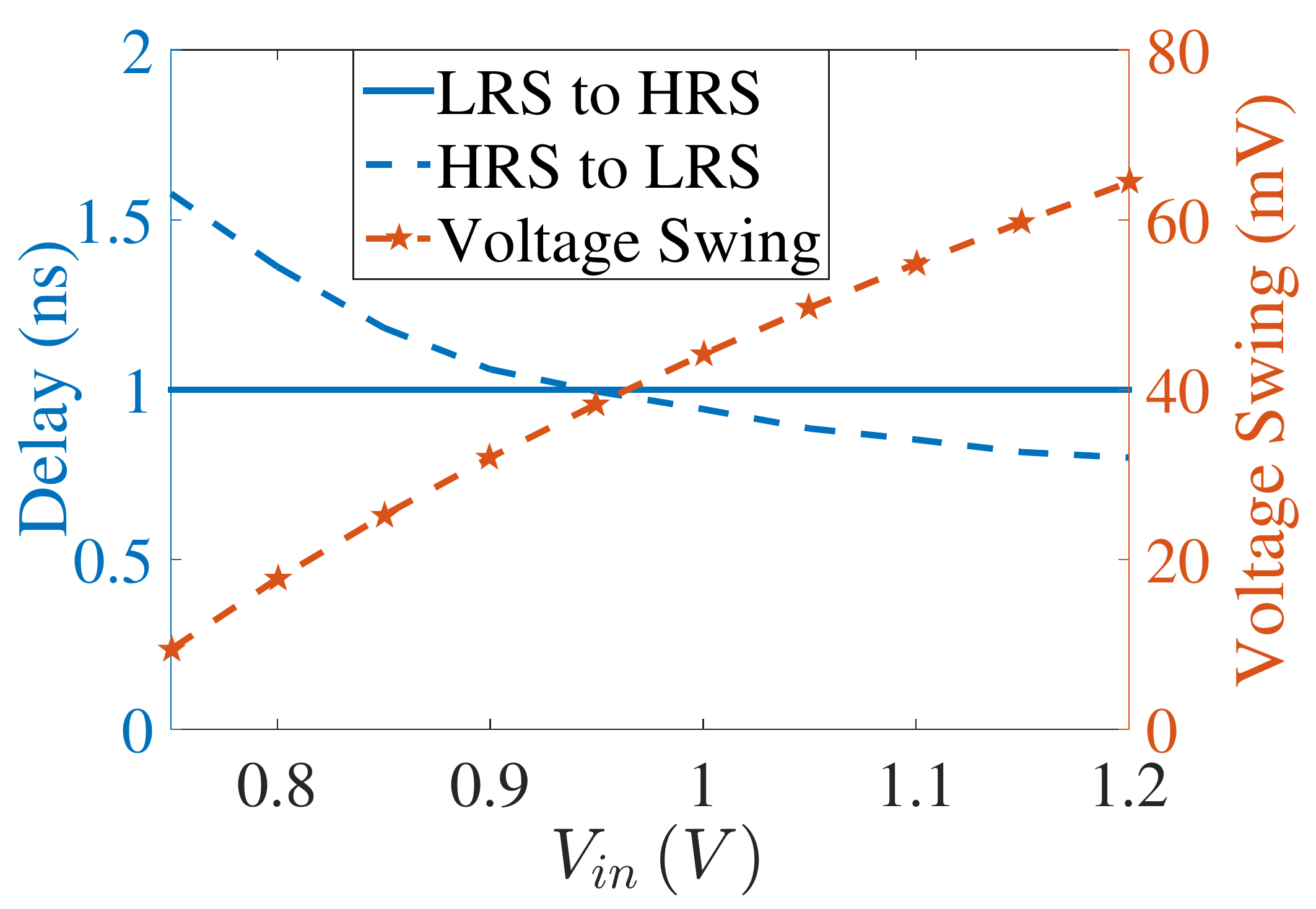}%
\label{Fig10b}}
\vspace{-0.1in}
\caption{Input comparator swing and delay versus (a) bias voltage with $V_{in}=1.2V$, (b) applied voltage with $V_{B}=0.7V.$ }
\label{Fig10}
\vspace{-0.25in}
\end{figure}

Figure \ref{Fig10a} shows the effect $V_B$ over the voltage swing and the delay. As previously discussed, the bias voltage should be greater than twice the transistor threshold voltage which is around $0.63V$. The output voltage swing exhibits critical curve with maximum point at around $V_B=0.75V$. Also, the delay for both reading scenarios (reading LRS then HRS and vice versa) is also shown. In our design, we set a practical target of $1ns$ for the delay. The best bias voltage is $0.7V$ to maximize the voltage swing. Another aspect that should be studied is the value of the input voltage. Figure \ref{Fig10b} shows the effect of changing $V_{in}$ with fixed $V_B=0.7$. 
The higher the input voltage, the more the voltage swing, the less delay and more power consumption as shown in Fig.\ref{PowerCurve}. 
 
{ $M_1-M_4$ transistors work as as a gain boosting circuit. Hence, the input impedance of the sensing circuit is high ($\approx 1M\Omega$). Due to the abrupt current changes while reading, $V_{in}$ is disturbed ($+20mv$ and $-35mV$ around $V_B$). The negative feedback of the gain boosting circuit works to recover $V_{in}$ to $V_B$. In our design, the loop recovers in $1ns$. Thus, the delay of the designed circuit is $1ns$. Practically, two phases are needed due to the latched comparator; (a) a reset phase where the output is set to $V_{DD}$ and the data is setup and (b) a latch phase where the data is latched and stored. Thus, another $1ns$ is needed to latch data for $50\%$ duty cycle clock. Per these parameters, the energy consumption of the readout circuit per bit is $7.6fJ$ for $500MHz$ clock frequency. }
\begin{table*}[!b]
\vspace{-0.1in}
\renewcommand{\arraystretch}{1.3}
\caption{Performance metric of reading a complete $N\times N$ array. These results are adopted from \cite{zidan2016single,naous2016pilot}. }
\vspace{-0.05in}
\label{T1}
\centering
\begin{tabular}{c c c c c c c }
\hline
\bfseries Technique & \bfseries Readout &\bfseries Throughput & \bfseries Locality& \bfseries Array                                & \bfseries Power\textsuperscript{2,*}& \bfseries FOM\textsuperscript{2}   \\
{}                            &   \bfseries Circuit    &                                  &\bfseries Needed& \bfseries Usage\textsuperscript{1} & \bfseries ($m W$)                           & \bfseries ($Tbits/W \mu m^2$)\\
\hline
Multistage\cite{vontobel2009writing}&ADC + Comp& $\frac{1}{6}$&No&1&7&0.04\\
Multiport\cite{zidan2014memristor}& ADC + Comp&$\frac{1}{3}$&No& $\frac{N-2}{N}$& 2.1 &0.265\\
Grounded Rows \& Cols\cite{manem2010design}&VG + Comp&1&No&1&4&0.4194  \\
Predefined Dummy Bits \cite{zidan2016single}&VG + Comp& $1$&Yes&$ \frac{N-1}{N} $&0.291&5.754\\
This work &VG + Comp&N/R&No&1&1.358 R&$633/R^2$\\
\hline
\end{tabular}
\\
\textsuperscript{1}Array Usage = number of data bits/total number of bits.\quad  \textsuperscript{2}{Power and FOM results are reported for 256Kb array.} \quad \textsuperscript{*} Without bias mismatch.
\vspace{-0.25in}
\end{table*}
\vspace{-0.15in}
\subsection{Figure of Merit}
In \cite{Fouda2017one}, a figure of merit (FOM) is defined for comparing different reading techniques taking into account important metrics such as throughput, and array usage. This FOM is defined as:
\begin{equation}
\!FOM=\frac{Throughput\times Array\,Usage}{Reading\,Power/cell \times \! pitch\,size}  (bit/W m^2)
\label{5.1}
\end{equation}
\noindent where the numerator reflects the array metrics; \textit{Throughput}, which is the number of read bits per cycle ({ bank size}), and \textit{Array Usage} which is the number of usable data bits divided by the total number of bits. The denumerator, reflects per bit the physical parameters, \textit{power per bit} and cell area. It is reported that the array density of memristor based selector-less crossbar arrays is approximately $640Gbit/cm^2$ where the feature size is $6.25nm$\cite{zidan2016single}. Table I shows a quantitative analysis for different reading techniques for a complete $N\times N$ array. We chose to compare with these four techniques which can accommodate high density crossbar arrays. This comparison illustrates the differences between prior work and the proposed approach in terms of power, throughput, array usage, and FOM.

The estimated area of the $512\times 512$ crossbar is around $40.96\mu m^2$ and the estimated area of the read decoder  is $4.26\mu m^2$ based on the technique published in \cite{balobas2017design,khan2015measuring} with two pre-decoding stages. To estimate the sensing circuitry area, a $128$ bit word is considered where the array is divided into 4 banks. The estimated area of 128 sensing circuit including the MUX is around $41.65\mu m^2$. The area of CMOS circuitry is estimated with respect to $5nm$ technology which is expected to be used with crossbar arrays. According to these estimated numbers, the entire CMOS circuitry can be placed under the crossbar array. And, the  static power dissipation is dominated by the sensing circuits which is around $285.7\mu W$ based on $1.92\mu W$ per bit.

{In this comparison, the exponential RRAM model is used with the aforementioned parameters values and feature size.} The main advantage of the proposed technique is the ability to read the entire row bits in one clock cycle which is vital in memories on the contrary with the other techniques which requires at least N clock cycles to read the entire row. However, the proposed technique consumes more power due to the nature of the reading technique. It is worth noting that other published works do not account for the readout circuitry, which should be accounted for in addition to any building blocks such as ADCs and comparators \cite{zidan2014memristor,vontobel2009writing}. The circuits proposed in this work can be used in other approaches as well such as with predefined dummy bits and grounded rows and columns \cite{manem2010design,zidan2016single}, which requires virtual ground and comparator.


\vspace{-0.1in}
\section{Bias Mismatch Effect}
{ In resistive memories, it is required to bias wordlines and bitlines to specific voltages based on the technique used. For example, in the proposed reading scheme, the unselected wordlines are connected to $V_B$, and all the bitlines are connected to $V_B$ through the sensing circuit. Thus, there would be a mismatch among the wordlines bias voltages themselves and with the bitlines bias voltages creating undesirable current due to PVT variation, wire parasitics and switch's resistance. This undesirable current is unavoidable and needs to be taken into consideration since it would limit the array size.    

The voltage mismatch is column independent and is not correlated to other columns. Hence, the number of resistive devices that are affected by the mismatch is $N-1$ devices. This mismatch is referred to as $\Delta V$. In case of linear switching devices, the current passing through each unwanted device is either $I_{LRS}=\Delta V/LRS $ or $I_{HRS}=\Delta V/HRS$. For equal probable states, the total unwanted current is $I_{unW}= \Delta V (0.5N/LRS+ (0.5 N-1)/HRS)$. Usually the ratio between $HRS$ and $LRS$ is $10^3$ or more, then the total unwanted current is approximated to $I_{unW}\approx 0.5N \Delta V/LRS$. However, the desired current for high/low resistance state is $I_{W}=(V_{DD}-V_B)/HRS$ or $I_{W}=(V_{DD}-V_B)/LRS$, respectively. Consequently, the extreme input current to the sensing circuit is $I_t=I_{W}\pm I_{unW}$ for high resistance and low resistance states, respectively. This sensed current affects the input voltage of the comparator $V_x$ directly, which should not exceed the noise margin of the comparator. For $10mV$ noise margin for the comparator, the maximum/minimum input sensed current, which are corresponding to $LRS$ and $HRS$, are $I_{max}=0.22\mu A$ and $I_{min}=0.195\mu A$, respectively. The maximum column width, $N$, is the minimum of $2*I_{max}*LRS/\Delta V$ or $2*I_{min}*LRS/\Delta V$ which is $195$ for $2mV$ bias mismatch.

On the other hand, in the case of nonlinear switching devices, the current passing through each unwanted device is $I_{LRS/HRS}=k_{on/off}\sinh(a\Delta V)$ for low/high resistance state. Since $\Delta V$ is very small in range of millivolts, the current can be approximated to $I_{LRS}\approx k_{on} a\Delta V$ or $I_{HRS}\approx k_{off} a\Delta V$. The total unwanted current is $I_{unW}= a\Delta V (0.5N k_{on}+ (0.5 N-1)k_{off})$. Usually the ratio between $k_{off}$ and $k_{on}$ is $10^3$ or more, then the total unwanted current is approximated to $I_{unW}\approx0.5N a\Delta V k_{on}$. However, the wanted current for high/low resistance states is $I_{W}=k_{off/on}\sinh(a(V_{DD}-V_B))$. Consequently, the extreme input current to the sensing circuit is $I_t=I_{W}\pm I_{unW}$ for high resistance and low resistance states, respectively. The maximum column width,$N$, is the minimum of $2*I_{max}/(\Delta V*a*k_{on})$ or $2*I_{min}/(\Delta V*a*k_{on})$ which is around $6500$ for $2mV$ bias mismatch. With nonlinear devices, the column width of the crossbar is highly increased due to the high resistance facing the mismatch voltage.   
\vspace{-0.15in}
\section{Conclusion and Future Perspective}
\vspace{-0.05in}
The proposed reading technique with the readout circuitry can read the entire row without using reference bits giving the maximum utilization of RRAM and the highest throughput for high speed applications. These advantages come with  more power consumption (4.7X more than \cite{zidan2016single}). According to the defined FOM taking into consideration, power consumption, array usage, effective array size, and throughput, the proposed technique is more than 100X better than \cite{zidan2016single} without banking. Moreover, according to the discussed studies for the sneak path immunity, power consumption and bias mismatch, the nonlinear devices are most recommended for high dense resistive memories. In addition, the proposed technique is compatible with the published writing techniques \cite{chen2015analysis} where switches are placed around the array to enable reading or writing since reading and witing can not be performed simultaneously in the same array.  

One of the features of the resistive memories is its ability to store multi-levels/multi-states such as ternary and quaternary data enabling higher higher radix processing units. The proposed technique can be applied for multilevel memories as well due to its ability to read the device resistance especially with nonlinear switching devices. However, the readout circuitry needs to modified to accommodate the multi-states and be able differentiate between them. This topic will be investigated in future research.     

Stack-ability of resistive memories is another feature enabling ultra dense memory arrays \cite{micheloni20163d}. The discussed readout technique alongside the circuitry can be used to read each crossbar layer by connecting the corresponding crossbar outputs together then to readout circuits. A level decoder is  needed to select the readout level. Using this configuration, only one row in a certain level is selected at a time where the other outputs result in zero output current. The stacked layers share the same reading circuitry which decreases the overhead of readout circuits. For instance, Xpoint memory is 2 layers resistive memories sharing the same bitlines and having two different wordlines
. The proposed readout circuitry can be connected to the bitlines and the wordlines are used to access the memory cells. The power density is one of the important aspects of any electronic circuit. By using the aforementioned technique, the  power density is approximately constant due to reading only one row per level at time. However, stack-ability might cause other reliability issues which limits the number of stacked layers, and is currently a subject of intensive research by the community.
}
\vspace{-0.2in}
\bibliographystyle{IEEEtran}
\bibliography{ref}

\end{document}